\documentclass[
 a4paper,
 10pt,
 twocolumn,
 superscriptaddress,
 amsmath,amssymb,
 aps,
 pra,
accepted=2020-10-05,
]{quantumarticle}

\pdfoutput=1

\usepackage{graphicx}
\usepackage{hyperref}

\usepackage{tikz}
\usepackage{float}

\begin{document}

\title{Transfer learning in hybrid classical-quantum neural networks}

\author{Andrea Mari}
\email{andreamari84@gmail.com}
\author{Thomas R. Bromley}
\author{Josh Izaac}
\author{Maria Schuld}
\author{Nathan Killoran}
\affiliation{Xanadu, 777 Bay Street, Toronto, Ontario, Canada.}


\begin{abstract}
We extend the concept of transfer learning, widely applied in modern machine learning algorithms, to the emerging context of hybrid neural networks composed of classical and quantum elements. We propose different implementations of hybrid transfer learning, but we focus mainly on the paradigm in which a pre-trained classical network is modified and augmented by a final variational quantum circuit. This approach is particularly attractive in the current era of intermediate-scale quantum technology since it allows to optimally pre-process high dimensional data (e.g., images) with any state-of-the-art classical network and to embed a select set of highly informative features into a quantum processor. We present several proof-of-concept examples of the convenient application of quantum transfer learning for image recognition and quantum state classification. We use the cross-platform software library PennyLane to experimentally test a high-resolution image classifier with two different quantum computers, respectively provided by IBM and Rigetti.
\end{abstract}

\maketitle


\section{Introduction}

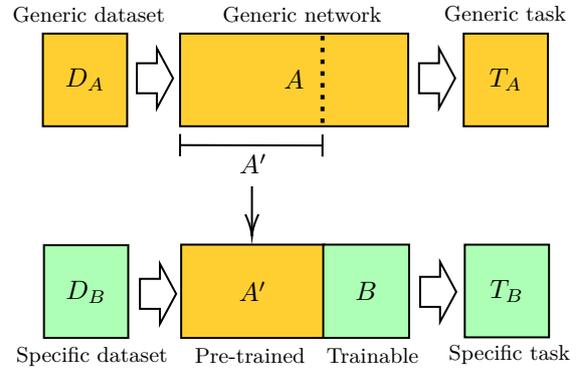
\begin{figure}
\vspace{1 em}

\tikzset{every picture/.style={line width=0.75pt}} 
\begin{tikzpicture}[x=0.75pt,y=0.75pt,yscale=-1,xscale=1]
\draw  [fill={rgb, 255:red, 255; green, 207; blue, 49 }  ,fill opacity=1 ] (31.98,66.29) -- (74.05,66.29) -- (74.05,112.9) -- (31.98,112.9) -- cycle ;
\draw  [fill={rgb, 255:red, 255; green, 207; blue, 49 }  ,fill opacity=1 ] (100.47,66.29) -- (214.35,66.29) -- (214.35,112.9) -- (100.47,112.9) -- cycle ;
\draw  [fill={rgb, 255:red, 255; green, 207; blue, 49 }  ,fill opacity=1 ] (242.04,66.29) -- (284.1,66.29) -- (284.1,112.9) -- (242.04,112.9) -- cycle ;
\draw   (78.9,81.2) -- (88.88,81.2) -- (88.88,73.65) -- (96.97,88.75) -- (88.88,103.86) -- (88.88,96.3) -- (78.9,96.3) -- cycle ;
\draw   (219.74,81.2) -- (229.71,81.2) -- (229.71,73.65) -- (237.8,88.75) -- (229.71,103.86) -- (229.71,96.3) -- (219.74,96.3) -- cycle ;
\draw  [fill={rgb, 255:red, 172; green, 255; blue, 180 }  ,fill opacity=1 ] (32.52,172.73) -- (74.59,172.73) -- (74.59,219.34) -- (32.52,219.34) -- cycle ;
\draw  [fill={rgb, 255:red, 172; green, 255; blue, 180 }  ,fill opacity=1 ] (242.58,172.46) -- (284.64,172.46) -- (284.64,219.07) -- (242.58,219.07) -- cycle ;
\draw   (80.44,189.64) -- (90.42,189.64) -- (90.42,182.09) -- (98.51,197.19) -- (90.42,212.3) -- (90.42,204.75) -- (80.44,204.75) -- cycle ;
\draw   (220.28,188.64) -- (230.25,188.64) -- (230.25,181.09) -- (238.34,196.19) -- (230.25,211.3) -- (230.25,203.75) -- (220.28,203.75) -- cycle ;
\draw    (136.34,143.11) -- (136.34,166.14) ;
\draw [shift={(136.34,168.14)}, rotate = 270] [color={rgb, 255:red, 0; green, 0; blue, 0 }  ][line width=0.75]    (10.93,-3.29) .. controls (6.95,-1.4) and (3.31,-0.3) .. (0,0) .. controls (3.31,0.3) and (6.95,1.4) .. (10.93,3.29)   ;

\draw    (100.47,122.39) -- (171.55,122.57) ;
\draw [shift={(171.55,122.57)}, rotate = 180.14] [color={rgb, 255:red, 0; green, 0; blue, 0 }  ][line width=0.75]    (0,5.59) -- (0,-5.59)   ;
\draw [shift={(100.47,122.39)}, rotate = 180.14] [color={rgb, 255:red, 0; green, 0; blue, 0 }  ][line width=0.75]    (0,5.59) -- (0,-5.59)   ;
\draw [line width=1.5]  [dash pattern={on 1.69pt off 2.76pt}]  (171.55,67.44) -- (171.55,113.93) ;

\draw  [fill={rgb, 255:red, 255; green, 207; blue, 49 }  ,fill opacity=1 ] (101.01,172.44) -- (172.47,172.44) -- (172.47,219.34) -- (101.01,219.34) -- cycle ;
\draw  [fill={rgb, 255:red, 172; green, 255; blue, 180 }  ,fill opacity=1 ] (171.93,172.44) -- (214.8,172.44) -- (214.8,219.34) -- (171.93,219.34) -- cycle ;

\draw (56.79,56.79) node [scale=0.9] [align=left] {{\small Generic dataset }};
\draw (161.41,56.79) node [scale=0.9] [align=left] {{\small Generic network}};
\draw (262.8,55.93) node [scale=0.9] [align=left] {{\small Generic task}};
\draw (58.41,228.84) node [scale=0.9] [align=left] {{\small Specific dataset }};
\draw (264.96,227.97) node [scale=0.9] [align=left] {{\small Specific task}};
\draw (135.43,227.97) node [scale=0.9] [align=left] {{\small Pre-trained}};
\draw (198.42,228.13) node [scale=0.9] [align=left] {{\small Trainable }};
\draw (53.01,89.59) node   {$D_{A}$};
\draw (53.55,196.04) node   {$D_{B}$};
\draw (157.41,89.59) node   {$A$};
\draw (136.74,195.89) node   {$A'$};
\draw (193.37,195.89) node   {$B$};
\draw (263.61,195.76) node   {$T_{B}$};
\draw (263.07,89.59) node   {$T_{A}$};
\draw (137.07,131.74) node   {$A'$};
\end{tikzpicture}
\caption{General representation of the {\it transfer learning} method, where each of the neural networks $A$ and $B$ can be either classical or quantum. Network $A$ is pre-trained on a dataset $D_A$ and for a task $T_A$. A reduced network $A'$, obtained by removing some of the final layers of $A$, is used as a fixed feature extractor. The second network $B$, usually much smaller than $A'$, is optimized on the specific dataset $D_B$ and for the specific task $T_B$. } \label{fig:transfer}
\end{figure}

Transfer learning is a typical example of an artificial intelligence technique that has been originally inspired by biological intelligence. It originates from the simple observation that the knowledge acquired in a specific context can be transferred to a different area. For example, when we learn a second language we do not start from scratch, but we make use of our previous linguistic knowledge. Sometimes transfer learning is the only way to approach complex cognitive tasks, e.g., before learning quantum mechanics it is advisable to first study linear algebra.
This general idea has been successfully applied also to design artificial neural networks \cite{pratt1993discriminability,pan2009survey,torrey2010transfer}. It has been shown \cite{raina2007self, yosinski2014transferable} that in many situations, instead of training a full network from scratch, it is more efficient to start from a pre-trained deep network and then optimize only some of the final layers for a particular task and dataset of interest (see Fig.~\ref{fig:transfer}).

The aim of this work is to investigate the potential of the transfer learning paradigm in the context of quantum machine learning \cite{biamonte2017quantum, schuld2015introduction, dunjko2016quantum}.  It is not in the scope of this study to make claims about the performance of quantum models with respect to classical models, but just to demonstrate that quantum transfer learning is feasible.
We focus on hybrid models \cite{farhi2018classification, schuld2019quantum, mcclean2016theory}, i.e., the scenario in which  quantum variational circuits \cite{peruzzo2014variational, schuld2018circuit, perdomo2018opportunities, mcclean2016theory, sim2019expressibility, killoran2018continuous} and classical neural networks can be jointly trained to accomplish hard computational tasks. In this setting, in addition to the standard {\it classical-to-classical} (CC) transfer learning strategy in which some pre-acquired knowledge is transferred between classical networks, three new variants of transfer learning naturally emerge: {\it classical to quantum} (CQ), {\it quantum to classical} (QC) and {\it quantum to quantum} (QQ).

In the current era of Noisy Intermediate-Scale Quantum (NISQ) devices \cite{preskill2018quantum}, CQ transfer learning is particularly appealing since it opens the possibility to classically pre-process large input samples (e.g., high resolution images) with any state-of-the-art deep neural network and to successively manipulate few but highly informative features with a variational quantum circuit. This scheme is quite convenient since it makes use of the power of quantum computers, combined with the successful and well-tested methods of classical machine learning.
On the other hand, QC and QQ transfer learning might also be very interesting approaches especially once large quantum computers will be available. In this case, fixed quantum circuits might be pre-trained as generic quantum feature extractors, mimicking well known classical models which are often used as pre-trained blocks: e.g., AlexNet \cite{krizhevsky2012imagenet}, ResNet \cite{he2016deep}, Inception \cite{szegedy2015going}, VGGNet \cite{simonyan2014very}, etc. (for image processing), or ULMFiT \cite{howard2018universal}, Transformer \cite{vaswani2017attention}, BERT \cite{devlin2018bert}, etc. (for natural language processing).
In summary, such classical state-of-the-art deep networks can either be {\it used} in CC and CQ transfer learning or {\it replaced} by quantum circuits in the QC and QQ variants of the same technique.

Up to now, the transfer learning approach has been largely unexplored in the quantum domain with the exception of a few interesting applications, for example, in modeling many-body quantum systems \cite{ch2017machine, huembeli2018identifying, zen2019transfer}, in the connection of a classical autoencoder to a quantum Boltzmann machine \cite{piat2018image} and in the initialization of variational quantum networks \cite{verdon2019learning}. With the present work we aim at developing a more general and systematic theory, specifically tailored to the emerging paradigms of variational quantum circuits and hybrid neural networks. 

For all the models theoretically proposed in this work, proof-of-principle examples of practical implementations are presented and numerically simulated. Moreover we also experimentally tested one of our models on physical quantum processors---{\it ibmqx4} by IBM  and {\it Aspen-4-4Q-A} by Rigetti---demonstrating the successful classification of high resolution images with a hybrid classical-quantum system. 
\section{Hybrid classical-quantum networks}

Before presenting the main ideas of this work, we begin by reviewing basic concepts of hybrid networks and introduce some notation.

\subsection{Classical neural networks}
A very successful model in classical machine learning is that of deep feed-forward neural networks \cite{goodfellow2016deep}.
The elementary block of a deep network is called a {\it layer} and  maps input vectors of $n_0$ real elements to output vectors of $n_1$ real elements. Its typical structure consists of an affine operation followed by a non-linear function $\varphi$ applied element-wise,
\begin{align}\label{eq:Clayer}
L_{n_0\rightarrow n_1}: \; {\bf x} \rightarrow {\bf y}=\varphi (W {\bf x} + {\bf b}).
\end{align}
Here, the subscript $n_0\rightarrow n_1$ indicates the number of input and output variables, $\bf x$ and $\bf y$ are the input and output vectors, $W$ is an $n_1 \times n_0$ matrix and $\bf b$ is a constant vector of $n_1$ elements.  The elements of $W$ and ${\bf b}$ are arbitrary real parameters (respectively known as {\it weights} and {\it baises}) which are supposed be {\it trained}, i.e., optimized for a particular task. The nonlinear function $\varphi$ is quite arbitrary but common choices are the {\it hyperbolic tangent} or the {\it rectified linear unit} defined as ${\rm ReLU}(x)= max(0,x)$.

A classical deep neural network is the concatenation of many layers, in which the output of the first is the input of the second and so on:
\begin{align}\label{eq:Cnet}
C = L_{n_{d-1}\rightarrow n_d} \circ \dots L_{n_1 \rightarrow n_2} \circ L_{n_0 \rightarrow n_1},
\end{align}
where  different layers have different weights.
Characteristic hyper-parameters of a deep network are its {\it depth} $d$ (number of layers) and the number of {\it features} (number of variables) for each layer, i.e., the sequence of integers $n_0,n_1\dots n_{d-1}$.

\subsection{Variational quantum circuits}
One of the possible quantum generalizations of feed-forward neural networks can be given in terms of {\it variational quantum circuits} \cite{farhi2018classification, schuld2019quantum, mcclean2016theory,peruzzo2014variational, schuld2018circuit, perdomo2018opportunities, mcclean2016theory, sim2019expressibility, killoran2018continuous}. Following the analogy with the classical case, one  can define a quantum layer as a unitary operation which can be physically realized by a low-depth variational circuit acting on the input state $|x\rangle$ of $n_q$ quantum subsystems (e.g., qubits or continuous variable modes) and producing the output state $|y\rangle$:
\begin{align}
\mathcal L:\; |x\rangle \rightarrow |y\rangle=U({\bf w}) |x\rangle,
\end{align}
where ${\bf w}$ is an array of classical variational parameters.
Examples of quantum layers could be: a sequence of single-qubit rotations followed by a fixed sequence of entangling gates \cite{schuld2018circuit, sim2019expressibility} or, for the case of optical modes, some active and passive Gaussian operations followed by single-mode non-Gaussian gates  \cite{killoran2018continuous}.
Notice that, differently from a classical layer, a quantum layer preserves the Hilbert-space dimension of the input states. This fact is due to the fundamental unitary nature of quantum mechanics and, as discussed at the end of this section, should be taken into account when designing quantum networks.

A variational quantum circuit of depth $q$ is a concatenation of many quantum layers, corresponding to the product of many unitaries parametrized by different weights:
\begin{align}
\mathcal Q =\mathcal L_q \circ \dots \mathcal L_2 \circ \mathcal L_1.
\end{align}
In order to inject classical data in a quantum network we need to embed a real vector ${\bf x}$ into a quantum state $|x\rangle$. This can also be done by a variational embedding layer depending on ${\bf x}$ and applied to some reference state (e.g., the vacuum or ground state),
\begin{align}
\mathcal E:\; {\bf x}  \rightarrow |x\rangle = E({\bf x}) |0\rangle.
\end{align}
Typical examples are single-qubit rotations or single-mode displacements parametrized by ${\bf x}$. Notice that, differently from $\mathcal L$, the embedding layer $\mathcal E$ is a map from a classical vector space to a quantum Hilbert space.

Conversely, the extraction of a classical output vector ${\bf y}$ from the quantum circuit can be obtained by measuring the expectation values of $n_q$ local observables ${\bf \hat y}= [\hat y_1, \hat y_2, \dots \hat y_{n_q}]$. We can define this process as a measurement layer, mapping a quantum state to a classical vector:
\begin{align}
\mathcal M: \;  |x\rangle  \rightarrow {\bf y} = \langle x |{\bf \hat y}|x \rangle .
\end{align}
Globally, the full quantum network including the initial embedding layer and the final measurement can be written as
\begin{align}\label{eq:Qnet}
Q = \mathcal M \circ \mathcal Q \circ \mathcal E.
\end{align}
The full network is a map from a classical vector space to a classical vector space depending on classical weights. Therefore, even though it may contain a quantum computation hidden in the quantum circuit, if considered from a global point of view, $Q$ is simply a black-box analogous to the classical deep network defined in Eq.~\eqref{eq:Cnet}.

However, especially when dealing with real NISQ devices, there are technical limitations and physical constraints which should be taken into account: while in the classical feed-forward network of Eq.~\eqref{eq:Cnet} we have complete freedom in the choice of the number of features for each layer; in the quantum network of Eq.~\eqref{eq:Qnet} all these numbers are often linked to the size of the physical system. For example, even if not strictly necessary, typical variational embedding layers encode each classical element of ${\bf x}$ into a single subsystem and so, in many practical situations, one has:
\begin{align}
{\rm \# inputs = \#\ subsystems = \#\ outputs}.
\end{align}
This common constraint of a variational quantum network could be overcome by: 
\begin{enumerate}
\item adding ancillary subsystems and discarding/measuring some of them in the middle of the circuit;
\item engineering more complex embedding and measuring layers;
\item adding pre-processing and post-processing classical layers. 
\end{enumerate} 
In this work, mainly because of its technical simplicity, we choose the third option and we formalize it through the notion of {\it dressed} quantum circuits introduced in the next subsection.

\subsection{Dressed quantum circuits}
In order to apply transfer learning at the classical-quantum interface, we need to connect classical neural networks to quantum variational circuits. Since in general the size of the classical and quantum networks can be very different, it is convenient to use a more flexible model of quantum circuits.

Let us consider the variational circuit defined in Eq.~\eqref{eq:Qnet} and based on $n_q$ subsystems. With the aim of adding some basic pre-processing and post-processing of the input and output data we place a classical layer at the beginning and at the end of the quantum network, obtaining what we might call a {\it dressed} quantum circuit:
\begin{align}\label{eq:dressed}
\tilde Q= L_{n_q \rightarrow n_{\rm out}} \circ Q \circ L_{n_{\rm in} \rightarrow n_q}.
\end{align}
where $L_{n \rightarrow n'}$ is given in Eq.~\eqref{eq:Clayer} and $Q$ is the associated {\it bare} quantum circuit defined in Eq.~\eqref{eq:Qnet}.
Differently from a complex hybrid network in which the computation is shared between cooperating classical and quantum processors, in this case the main computation is performed by the quantum circuit $Q$, while the classical layers are mainly responsible for the data embedding and readout. A similar hybrid model was applied to a generative quantum Helmholtz machine in Ref.~\cite{benedetti2018quantum} and to a variational circuit in Ref.~\cite{adhikary2020supervised}.

We can say that from a hardware point of view a dressed quantum circuit is almost equivalent to a bare
one. On the other hand, it has two important advantages:
\begin{enumerate}
\item the two classical layers can be trained to optimally perform the embedding of the input data and the post-processing of the measurement results;
\item the number of input and output variables are independent from the number of subsystems, allowing for  flexible connections to other classical or quantum networks.
\end{enumerate} 

Even if our main motivation for introducing the notion of dressed quantum circuits is a smoother implementation of transfer learning schemes, this is also a quite powerful machine learning model in itself and constitutes a non-trivial contribution of this work. In the Examples section, a dressed quantum circuit is successfully applied to the classification of a non-linear benchmark dataset (2D spirals).

\section{Transfer learning}\label{sec:transfer}

In this section we discuss the main topic of this work, i.e., the idea of transferring some pre-acquired ``knowledge" between two networks, say from network $A$ to network $B$, where each of them could be either classical or quantum. 

As discussed in the previous section, if considered as a black box,  the global structure of a quantum variational circuit is similar to that of a classical network (see Eqs.~\eqref{eq:Qnet},   \eqref{eq:dressed} and~\eqref{eq:Cnet}).
For this reason, we are going to define the transfer learning scheme in terms of two generic networks $A$ and $B$, independently from their classical or quantum physical nature.\\

\noindent{\it Generic transfer learning scheme (see Fig.~\ref{fig:transfer}):}

\begin{enumerate}
\item Take a network $A$ that has been pre-trained on a dataset $D_A$ and for a given task $T_A$.
\item Remove some of the final layers. In this way, the resulting truncated network $A'$ can be used as a feature extractor.
\item Connect a new trainable network $B$ at the end of the pre-trained network $A'$.
\item Keep the weights of $A'$ constant, and train the final block $B$ with a new dataset $D_B$ and for a new task of interest $T_B$.
\end{enumerate}
Following the common convention used in classical machine learning  \cite{pratt1993discriminability,pan2009survey,torrey2010transfer,raina2007self, yosinski2014transferable}, all situations in which there is a change of dataset $D_B \neq D_A$ and/or a change of the final task $T_B \neq T_A$ can be identified as {\it transfer learning} methods. The general intuition behind this training approach is that, even if $A$ has been optimized for a specific problem it can still act as a convenient feature extractor also for a different problem. 

The concept of feature extraction is much more general than what is actually needed in this work (see, e.g., Ref.~\cite{bengio2013representation} for a good introduction to this topic).
In the context of transfer learning, the truncated pre-trained network $A'$ is interpreted as a feature extractor because, after removing some of the final layers of $A$ (step 2), $A'$ can be used to produce features which are not problem-specific. The motivation for truncating the final layers of $A$ is that the final activations of a network are usually more tuned to the specific problem, while intermediate features are more generic and so more suitable for transfer learning.  For example, in a convolutional neural network, a feature extractor can be obtained by removing the final dense layers while preserving all the pre-trained convolution layers.

In our hybrid setting, the fact that the networks $A$ and $B$ can be either classical or quantum gives rise to a rich variety of hybrid transfer learning models summarized in Table \ref{tab:models}.
\begin{table}[t]
\begin{tabular}{l|l|l}
\bf A       & \bf B 		& \bf Transf. learn. scheme        			\\ 
\hline
\hline
Classical 	& Classical 	& CC (\cite{pratt1993discriminability,
pan2009survey,torrey2010transfer,raina2007self, yosinski2014transferable}) 	\\ 
Classical 	& Quantum   	& CQ (Examples 2 and 3)		 			 	\\ 
Quantum $\ $& Classical 	& QC (Example 4)								\\ 
Quantum   	& Quantum   	& QQ (Example 5) 								\\ 
\end{tabular}
\caption{Transfer learning schemes in hybrid classical-quantum networks. The corresponding proof-of-principle examples are indicated in the table and presented in Section \ref{sec:examples}.} \label{tab:models}
\end{table}
For the reader familiar with quantum communication theory, this kind of classification might look similar to that of hybrid channels in which information can be exchanged between quantum and classical systems. Here however there is a fundamental difference: what is actually transferred in this case is not raw information but some more structured and organized {\it learned representations}. We expect that the problem of
transferring structured knowledge between systems governed by different physical laws (classical/quantum) could stimulate many interesting foundational and philosophical questions. The aim of the present work is however much more pragmatic and consists of studying practical applications of this idea.

\subsection{Classical to quantum transfer learning}

As discussed in the introduction, the CQ transfer learning approach is perhaps the most appealing  one in the current technological era of NISQ devices. Indeed today we are in a situation in which intermediate-scale quantum computers are approaching the quantum supremacy milestone \cite{harrow2017quantum, arute2019quantum} and, at the same time, we have at our disposal the very successful and well-tested tools of classical deep learning. The latter are  universally recognized as the best-performing machine learning algorithms,  especially for image and text processing. 

In this classical field, transfer learning is already a very common approach, thanks to the large zoo of pre-trained deep networks which are publicly available \cite{canziani2016analysis}. CQ transfer learning consists of using exactly those classical pre-trained models as feature extractors and then post-processing such features on a quantum computer; for example by using them as input variables for the dressed quantum circuit model introduced in Eq.~\eqref{eq:dressed}.
This hybrid approach is very convenient for processing high-resolution images since, in this configuration, a quantum computer is applied only to a fairly limited number of abstract features, which is much more feasible compared to embedding millions of raw pixels in a quantum system. We would like to mention that also other alternative approaches for dealing with large images have been recently proposed \cite{piat2018image, henderson2019quanvolutional, shiba2019convolution, liu2019machine}.

We applied our model for the task of image classification in several numerical examples and we also tested the algorithm with two real quantum computers provided by IBM and Rigetti. All the details about the technical implementation and the associated results are reported in the next Section, Examples 2 and 3.

\subsection{Quantum to classical transfer learning}

By switching the roles of the classical and quantum networks, one can also obtain the QC variant of transfer learning.
In this case a pre-trained quantum system behaves as a kind of feature extractor, i.e., a device performing a (potentially classically intractable) computation resulting in an output vector of numerical values associated to the input. As a second step, a classical network is used to further process the extracted features for the specific problem of interest. 
This scheme can be very useful in two important situations: $(i)$ if the dataset consists of
quantum states (e.g., in a state classification problem), $(ii)$ if we have at our disposal a very good quantum computer which outperforms current classical feature extractors at some task. 

For case $(i)$, one can imagine a situation in which a single instance of a variational quantum circuit is first pre-trained and then used as a kind of {\it multipurpose measurement device}. Indeed one could make many different experimental analyses by simply letting input quantum systems pass through the same fixed circuit and applying different classical machine learning algorithms to the associated measured variables.

For case $(ii)$ instead, one can envisage a multi-party scenario in which many classical clients $B$ can independently send samples of their specific datasets to a common quantum server $A$ which is pre-trained to extract generic features by performing a fixed quantum computation. Server $A$ can send back the resulting features to the classical clients $B$, which can now locally train their specific machine learning models on pre-processed data.

Given the current status of quantum technology, case $(ii)$ is likely beyond a near-term implementation. On the other hand, case $(i)$ could already represent a realistic scenario with current technology.\\

In Example 4 of the next Section, we present a proof-of-concept example in which a pre-trained quantum network introduced in Ref.\ \cite{killoran2018continuous} is combined with a classical post-processing network for solving a quantum state classification problem.

\subsection{Quantum to quantum transfer learning}

The last possibility is the QQ transfer learning scheme, where the same technique is applied in a fully quantum mechanical fashion. In this case a quantum network $A$ is pre-trained for a generic task and dataset. Successively, some of the final quantum layers are removed, and replaced by a trainable quantum network $B$ which will be optimized for a specific problem. The main difference from the previous cases is that, since the process is fully quantum without intermediate measurements, features are implicitly transferred in the form of a quantum state, allowing for coherent superpositions. 

The main motivation for applying a QQ transfer learning scheme is to reduce the total training time: instead of training a large variational quantum circuit, it is more efficient to initialize it with some pre-trained weights and then optimize only a couple of final layers. From a physical point of view, such optimization of the final layers could be interpreted as a change of the measurement basis which is tuned to the specific problem of interest.

If compared with classical computers, current NISQ devices are not only noisy and small: they are also relatively slow.
Training a quantum circuit might take a long time since it requires taking many measurement shots (i.e., performing a large number of actual quantum experiments) for each optimization step (e.g., for computing the gradient). Therefore any approach which can reduce the total training time, as for example the QQ transfer learning scheme, could be very helpful.

In Example 5 of the next Section, we trained a quantum state classifier by following a QQ transfer learning approach.

\section{Examples}
\label{sec:examples}

The Python code related to the examples presented in this section is available at~\cite{transfer_learning_code}.

\subsection*{Example 1 -  A 2D classifier based on a dressed quantum circuit}

This first example demonstrates the {\it dressed} quantum circuit model introduced in Eq.~\eqref{eq:dressed}.
\begin{figure}[ht]
\includegraphics[width=0.45\textwidth]{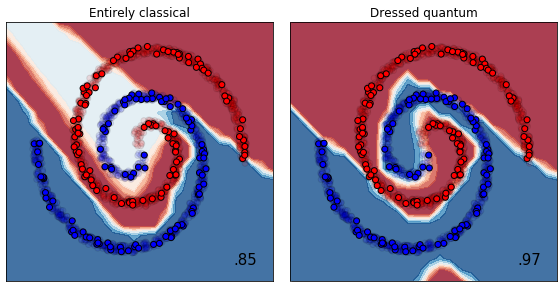}
\caption{Classification based on a classical network (left) and a dressed quantum circuit (right) of the same dataset consisting of two set of points (blue and red) organized in two concentric spirals. Training points are pale-colored while test points are sharp-colored. The model decision function is evaluated for the whole 2D plane, determining the blue and red regions which should ideally match the color of the data points. The test accuracy of each model is reported in the bottom-right of the corresponding plot.} \label{fig:spirals}
\end{figure}
We consider a typical benchmark dataset consisting of two classes of points (blue and red) organized in two concentric spirals as shown in Fig.~\ref{fig:spirals}. Each point is characterized by two real coordinates and we  assume to have at our disposal a quantum processor of 4 qubits. Since we have two real coordinates as input and two real variables as output (one-hot encoding the blue and red classes), we use the following model of a dressed quantum circuit:
\begin{align}\label{eq:dressed_242}
\tilde Q = L_{4 \rightarrow 2} \circ Q \circ L_{2 \rightarrow 4},
\end{align}
where $L_{2 \rightarrow 4}$ represents a classical layer having the structure of Eq.~\eqref{eq:Clayer} with $\varphi={\rm tanh}$, $Q$ is a (bare)  variational quantum circuit, and $L_{4 \rightarrow 2}$ is a linear classical layer without activation i.e., with $\varphi({\bf y})={\bf y}$.
The structure of the variational circuit is $Q=\mathcal M \circ \mathcal Q \circ \mathcal E$ as in Eq.~\eqref{eq:Qnet}. The chosen embedding map prepares each qubit in a balanced superposition of $|0 \rangle$ and $|1\rangle$ and then performs a rotation around the $y$ axis of the Bloch sphere parametrized by a   classical vector $\bf x$:
\begin{align} \label{eq:E_24}
\mathcal E({\bf x})= \bigotimes_{k=1}^4 \Big( R_y(x_k \pi/2) H \Big)|0000\rangle,
\end{align}
where $H$ is the single-qubit Hadamard gate.
The trainable circuit is composed of 5 variational layers $\mathcal Q= \mathcal L_5 \circ \mathcal L_4 \circ \mathcal L_3 \circ \mathcal L_2 \circ \mathcal L_1 $, where
\begin{align} \label{eq:L_44}
\mathcal L({\bf w}):\; |x\rangle \rightarrow |y\rangle = K \bigotimes_{k=1}^4  R_y(w_k) |x\rangle,
\end{align}
and $K$ is an entangling unitary operation made of three controlled NOT gates:
\begin{align} \label{eq:K_44}
K= (CNOT\otimes \mathbb I_{3,4})( \mathbb I_{1,2} \otimes CNOT)(\mathbb I_1 \otimes CNOT\otimes \mathbb I_4).
\end{align}

Finally, the measurement layer is simply given by the expectation value of the  $Z={\rm diag}(1,-1)$ Pauli matrix, locally estimated for each qubit:
\begin{equation} \label{eq:M_44}
\mathcal M(|y\rangle) = {\bf y}= \left[
\begin{array}{c}
    \langle y| Z\otimes \mathbb I\otimes \mathbb I\otimes \mathbb I |y\rangle \\
    \langle y| \mathbb I\otimes Z\otimes \mathbb I\otimes \mathbb I |y\rangle \\
    \langle y| \mathbb I\otimes \mathbb I\otimes Z\otimes \mathbb I |y\rangle \\
    \langle y| \mathbb I\otimes \mathbb I\otimes \mathbb I\otimes Z |y\rangle 
\end{array}
\right]
\end{equation}
Given an input point of coordinates ${\bf x}=(x_1,x_2)$, the classification is done according to ${\rm argmax} ( {\bf y})$, where ${\bf y}=(y_1,y_2)$ is the output of the dressed quantum circuit \eqref{eq:dressed_242}.

For training and testing the model, the dataset has been divided into 2000 training points (pale-colored in Fig.~\ref{fig:spirals}) and 200 test points (sharp-colored in Fig.~\ref{fig:spirals}). As typical in classification problems, the cross entropy (implicitly preceded by a LogSoftMax layer) was used as a loss function. 
The LogSoftMax layer 
acts as $y_j \rightarrow {\hat p_j} = e^{y_j}/(e^{y_1} + e^{y_2})$, such that $\{\hat p_j\}$ is a valid probability distribution for $j=1,2$. The predicted distribution $\{\hat p_j\}$ should approximate the true  one $\{p_j\}$, which in this example is given by: $\{p_1=1,\, p_2=0\}$, if $\bf x$ belongs to the red class; $\{p_1=0,\, p_2=1\}$, if $\bf x$ belongs to the blue class. The cross entropy loss function is defined as:

\begin{equation} \label{eq:cross_entropy}
J =  - \mathbb{E}\sum_j p_j \log(\hat p_j)= - \mathbb{E}\sum_j p_j [y_j - \log(\sum_k e^{y_k})], 
\end{equation}
where the average $\mathbb{E}$ is with respect to the train/test datasets. 
The right-hand-side of Eq.~\eqref{eq:cross_entropy} was minimized via the Adam optimizer \cite{kingma2014adam}. 
A total number of 1000 training iterations were performed, each of them with a batch size of 10 input samples. The numerical simulation was done through the PennyLane software platform \cite{bergholm2018pennylane}.

The results are reported in Fig.~\ref{fig:spirals}, where the dressed quantum network is also compared with an entirely classical counterpart in which the quantum circuit is replaced by a classical layer, i.e., $C=L_{4 \rightarrow 2} \circ L_{4 \rightarrow 4} \circ L_{2 \rightarrow 4}$. The corresponding accuracy, i.e., the fraction of test points correctly classified, is $0.97$ for the dressed quantum circuit and $0.85$ for the classical network.

The presented results suggest that a dressed quantum circuit is a very flexible quantum machine learning  model which is capable of classifying highly non-linear datasets. We would like to remark that the classical counter-part has been presented just as a qualitative benchmark: even if in this particular example the quantum model outperforms the classical one, any general and rigorous comparison would require a much more complex and detailed analysis which is beyond the aim of this work.
It is also worth to mention that the training time was significantly longer for the (simulated) quantum model ( $\approx 2$ sec) with respect to the classical model ($\approx 0.01$ sec).

\subsection*{Example 2 -  CQ transfer learning for image classification (ants / bees)}

In this second example we apply the classical-to-quantum transfer learning scheme for solving an image classification problem. We first numerically trained and tested the model, using PennyLane with the PyTorch \cite{paszke2017automatic} interface. Successively, we have also run it on two real quantum devices  provided by {\it IBM} and {\it Rigetti}. To our knowledge, this is the first time that high resolution images have been classified with a hybrid classical-quantum system.

Our example is a quantum model inspired by the official PyTorch tutorial on classical transfer learning \cite{chilamkurthy2017tutorial}.
The model can be defined in terms of the general CQ scheme proposed in Section \ref{sec:transfer} and represented in Fig.~\ref{fig:transfer}, with the following specific settings:

\begin{description}
\item[$D_A$ =] ImageNet: a public image dataset with 1000 classes \cite{deng2009imagenet}.
\item[$A$ =] RestNet18: a pre-trained residual neural network introduced by Microsoft in 2016 \cite{he2016deep}.
\item[$T_A$ =] Classification (1000 labels).
\item[$A'$ =] RestNet18 without the final linear layer, obtaining a pre-trained extractor of 512 features.
\item[$D_B$ =] Images of two classes: {\it ants} and {\it bees} ({\it Hymenoptera} subset of ImageNet), separated into a training set of 245 images and a testing set of 153 images.
\item[$B$ =] $\tilde Q = L_{4 \rightarrow 2} \circ Q \circ L_{512 \rightarrow 4}$ : i.e., a 4-qubit dressed quantum circuit \eqref{eq:dressed} with 512 input features and 2 real outputs.
\item[$T_B$ =] Classification (2 labels).
\end{description}

\begin{figure}[t]
\includegraphics[width=0.38 \textwidth]{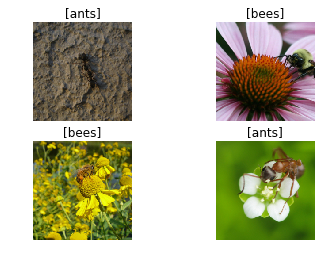}
\caption{Random batch of 4 images sampled form the test dataset $D_B$ and classified by our classical-quantum model (numerically simulated). Predictions are reported in square brackets above each image.}\label{fig:antsbees}
\end{figure}

The bare variational circuit is essentially the same as the one used in the previous example (see Eqs.~(\ref{eq:E_24},\ref{eq:L_44},\ref{eq:K_44},\ref{eq:M_44})), with the only difference that in this case the quantum depth is set to $6$.
The cross entropy is used as a loss function and minimized via the Adam optimizer \cite{kingma2014adam}.
We trained the variational parameters of the model for $30$ epochs over the training dataset, with a batch size of $4$ and an initial learning rate of $\eta=0.0004$, which was successively reduced by a factor of $0.1$ every 10 epochs.
After each epoch, the model was validated with respect to the test dataset, obtaining a maximum accuracy of $0.967$. A visual representation of a random batch of images sampled from the test dataset and the corresponding predictions is given in Fig.~\ref{fig:antsbees}. 

\begin{figure}[t]
\includegraphics[width=0.5 \textwidth]{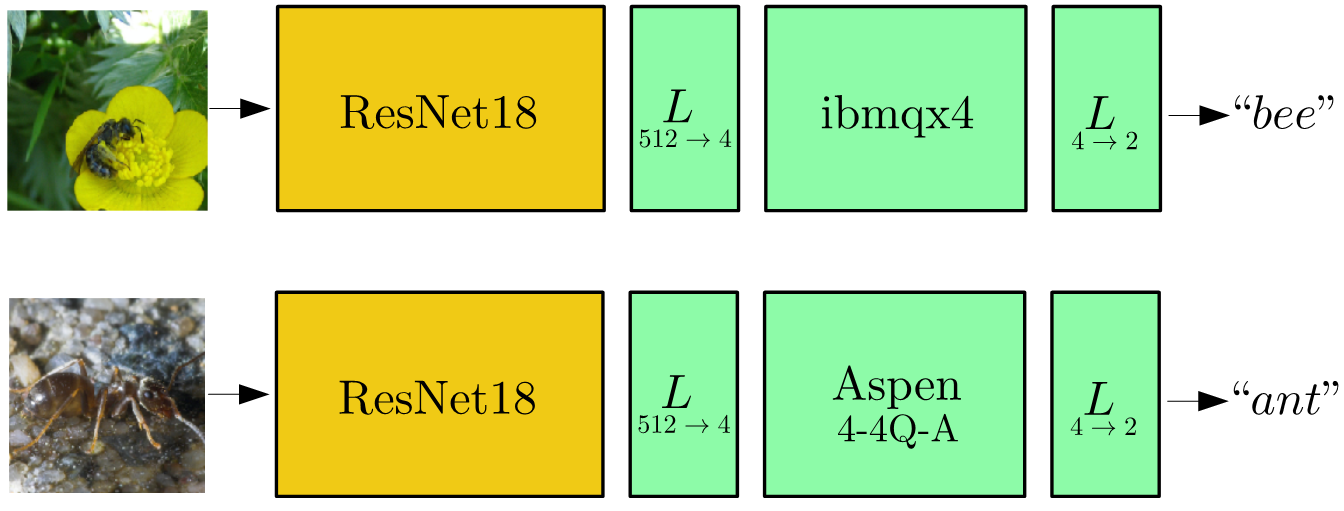}
\caption{Two high-resolution images sampled from the dataset $D_B$ and experimentally classified with two different quantum processors: {\it ibmqx4} by IBM (first line) and {\it Aspen-4-4Q-A} by Rigetti (second line). In both cases the same pre-trained classical network (ResNet18 by Microsoft \cite{he2016deep}) was used to pre-process the input image, extracting 512 highly informative features. The rest of the computation was performed by a trainable variational quantum circuit  ``dressed" by two classical encoding and decoding layers, as described in Eq.~\eqref{eq:dressed}.  }
\label{fig:realQPUs}
\end{figure}

We also tested the model (with the same pre-trained parameters), on two different real quantum computers: the {\it ibmqx4} processor by IBM and the {\it Aspen-4-4Q-A} processor by Rigetti (see Fig.~\ref{fig:realQPUs}). The corresponding classification accuracies, evaluated on the same test dataset, are reported in Table \ref{tab:QPU}. For the simulation, the exact expectation values were numerically evaluated. On the other hand, for both experiments performed on real quantum processors, we averaged 1024 measurement shots for each expectation value. 

We did not apply any particular strategy to reduce the execution time of the hybrid model since it was not relevant for our aims. However, for the sake of completeness, we report that the time necessary to test the model over the full dataset was approximately:  $\approx 26$~sec when using {\it ibmqx4} with a free account, $\approx 7$~sec when using {\it Aspen-4-4Q-A}, and $\approx 2$~sec when using the default PennyLane simulator on a laptop. 
As a qualitative reference, we also report that time corresponding to the classical network ResNet18 (without any quantum layer) was $\approx 1$~sec.   
 
\begin{table}[t]
\begin{center}
\begin{tabular}{l|l}
\bf QPU     & \bf Accuracy \\ 		
\hline
\hline
Simulator 	    & $0.967$ \\
ibmqx4      	& $0.95$ \\
Aspen-4-4Q-A	& $0.80$ 
\end{tabular}
\end{center}
\caption{Image classification accuracy associated to different quantum processing units (QPU): exact simulator, {\it ibmqx4} (IBM) and {\it Aspen-4-4Q-A} (Rigetti). } \label{tab:QPU}
\end{table}

Our results demonstrate the promising potential of the CQ transfer learning scheme applied to current NISQ devices, especially in the context of high-resolution image processing.

\subsection*{Example 3 -  CQ transfer learning for image classification (CIFAR)}

\begin{figure}[t]
{\bf (a)} \hspace{4 cm} {\bf (b)}    \\                 
\includegraphics[width=0.22 \textwidth]{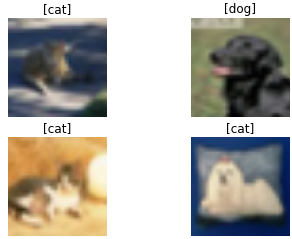}
\hspace{0.5 cm}
\includegraphics[width=0.22 \textwidth]{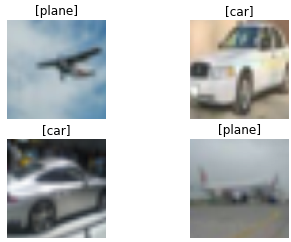}
\caption{Random batches of 4 images sampled from the CIFAR-10 test dataset restricted to the classes of {\it cats} and {\it dogs} (a), and to the classes of {\it planes} and {\it cars} (b). In both cases the binary classification problem is solved by our hybrid classical-quantum model (numerically simulated). Predictions are reported in square brackets above each image.}\label{fig:CIFAR}
\end{figure}

We now apply the same CQ transfer learning scheme of the previous example but with a different dataset $D_B$. Instead of classifying images of {\it ants} and {\it bees}, we use the standard CIFAR-10 dataset \cite{krizhevsky2009learning} restricted to the classes of {\it cats} and {\it dogs}. Successively we also repeat again the training and testing phases with the CIFAR-10 dataset restricted to the classes of {\it planes} and {\it cars} (see Fig.~\ref{fig:CIFAR}).

We remark that, in both cases, the feature extractor ResNet18 is pre-trained on ImageNet. Despite CIFAR-10 and ImageNet being quite different datasets (they also have very different resolutions), the CQ transfer learning method achieves nonetheless relatively good results.

\begin{table}[H]
\begin{center}
\begin{tabular}{l|l|l|l}
                    &\bf ants/          & \bf dogs/     & \bf planes/  \\ 		
                    &\bf bees           & \bf cats      & \bf cars     \\ 		
\hline
\hline
Quantum depth       & 6                 & 5             & 4                 \\
Number of epochs 	& 30                & 3             & 3                 \\
Batch size      	& 4                 & 8             & 8                 \\
Learning rate	    & 0.0004            & 0.001         & 0.0007            \\
\hline
Accuracy    & 0.976             & 0.8270        & 0.9605    
\end{tabular}
\end{center}
\caption{Hyper-parameters used for the classification of different image datasets $D_B$. The last line reports the corresponding accuracy achieved by our model (numerically simulated).} \label{tab:paramsacc}
\end{table}

The hyper-parameters used for all the different datasets (including the previous example) are summarized in Table \ref{tab:paramsacc}. The corresponding test accuracies are also reported in Table \ref{tab:paramsacc}, while some of the predictions for random samples of the restricted CIFAR datasets are visualized in Fig.~\ref{fig:CIFAR}.

\subsection*{Example 4 -  QC transfer learning for quantum state classification}

Quantum to classical (QC) transfer learning consists of using a pre-trained quantum circuit as a feature extractor and in post-processing its output variables with a classical neural network. In this case only the final classical part will be trained to the specific problem of interest.

\begin{figure}[t]
\includegraphics[width=0.45 \textwidth]{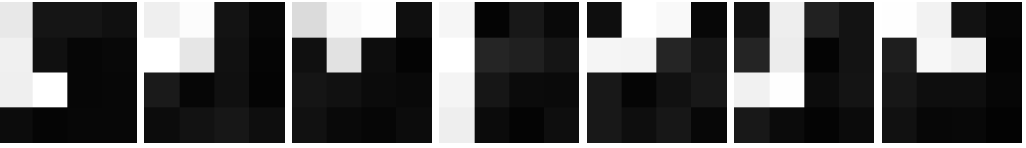} 
\caption{Images of the 7 different tetrominos encoded in the photon number probability distribution of two optical modes, after projecting on the subspace of up to $3$ photons. These images are extracted from Fig.~10 of Ref.\ \cite{killoran2018continuous}.}\label{fig:tetris}
\end{figure}

The starting point of our example is the pre-trained continuous-variable quantum network presented in Ref.\ \cite{killoran2018continuous}, Section IV.D, Experiment C. The original aim of this network was to encode $7$ different $4 \times 4$ images, representing the (L,O,T,I,S,J,Z) tetrominos (popularized by the video game {\it Tetris} \cite{wikipedia2019tetris}), in the Fock basis of two-mode quantum states. The expected input of the quantum network is one of the following combinations of two-mode coherent states:
\begin{align}
|\varphi_1\rangle &= |\alpha\rangle|\alpha\rangle \label{eq:tetris} \\
|\varphi_2\rangle &= |-\alpha\rangle|\alpha\rangle \nonumber \\
|\varphi_3\rangle &= |\alpha\rangle|-\alpha\rangle \nonumber \\
|\varphi_4\rangle &= |-\alpha\rangle|-\alpha\rangle \nonumber \\
|\varphi_5\rangle &= |i \alpha\rangle| i\alpha\rangle \nonumber \\
|\varphi_6\rangle &= |-i \alpha\rangle|i \alpha\rangle \nonumber \\
|\varphi_7\rangle &= |i \alpha\rangle|-i \alpha\rangle  \nonumber 
\end{align}
where the parameter $\alpha=1.4$ is a fixed constant. 
In Ref.\ \cite{killoran2018continuous} the network was successfully trained to generate an optimal unitary operation $|\tilde{\varphi}_j\rangle=U|\varphi_j\rangle$, such that the probability of finding $i$ photons in the first mode and $j$ photons in the second mode is proportional to the amplitude of the image pixel $(i,j)$. More precisely, the network was trained to reproduce the tetromino images after projecting the quantum state on the subspace of up to 3 photons (see Fig.~\ref{fig:tetris}).

For the purposes of our example, we now assume that the previous $7$ input states \eqref{eq:tetris} are subject to random Gaussian displacements in phase space:

\begin{align} \label{eq:noise}
|\varphi_j & \rangle   \xrightarrow{\text{random displacement}}  \hat D (\delta\alpha_1,\delta\alpha_2)|\varphi_j \rangle,
\end{align}
where $ \hat D(\alpha_1,\alpha_2)$ is a two-mode displacement operator \cite{weedbrook2012gaussian}, the values of the complex displacements $\delta\alpha_1$ and $\delta\alpha_2$ are sampled from a symmetric Gaussian distribution with zero mean and quadrature  variance $N \ge0$, and $j \in \{1,2,3,4,5,6,7 \}$ is the label associated to the input states \eqref{eq:tetris}.
The noise is similar to a Gaussian additive channel \cite{weedbrook2012gaussian}; however, for simplifying the numerical simulation, here we assume that the unknown displacements remain constant during the estimation of expectation values. Physically, this situation might represent a slow phase-space drift of the input light mode.

We also assume that, differently from the original image encoding problem studied in Ref.\ \cite{killoran2018continuous}, our new task is to classify the noisy input states. In other words, the network should take the states defined in \eqref{eq:noise} as inputs, and should ideally produce the correct label $j \in \{1,2,3,4,5,6,7 \}$ as output. 
In order to tackle this problem, we apply a QC transfer learning approach: we pre-process our random input states with the quantum network of Ref. \cite{killoran2018continuous} and we consider the corresponding images as features which we are going to post-process with a classical layer to predict the state label $j$. In simple terms, the QC transfer learning method allows us to convert a quantum state classification problem into an image classification problem.

Also in this case we can summarize the transfer learning scheme according to the notation introduced in  Section \ref{sec:transfer} and represented in Fig.~\ref{fig:transfer}:
\begin{description}
\item[$D_A$ =] $7$ two-mode coherent states defined in Eq.~\eqref{eq:tetris}.
\item[$A$ =] Photonic neural network introduced in Ref.\ \cite{killoran2018continuous}, consisting of an encoding layer, $25$ variational layers, and a final (Fock) measurement layer.
\item[$T_A$ =] Fock basis encoding of tetrominos images (see Fig.~\ref{fig:tetris}).
\item[$A'$ =] Pre-trained network $A$, truncated up to a quantum depth of $q=15$ variational layers.
\item[$D_B$ =] Same states of the original dataset $D_A$ but subject to random phase-space displacements as described in Eq.~\eqref{eq:noise}.
\item[$B$ =] $L_{16 \rightarrow 7}$ : i.e., a classical linear layer having the structure of Eq.~\eqref{eq:Clayer}, without activation ($\varphi({\bf y})={\bf y}$).
\end{description}

Also in this case we used the Adam optimizer \cite{kingma2014adam} to minimize a cross-entropy loss function associated to our classification problem. For each optimization step we sampled independent random displacements with variance $N=0.6$ that we applied to a batch of $7$ states defined in Eq.~\eqref{eq:tetris}. We optimized the model over 1000 training batches with a learning rate of $\eta=0.01$, obtaining a classification accuracy of $0.803$. The numerical simulation was performed with the Strawberry Fields software platform \cite{killoran2019strawberry}, combined with the TensorFlow \cite{abadi2016tensorflow} optimization back-end.

A summary of the hyper-parameters and of the corresponding accuracy is given in Table \ref{tab:tetris}.

\begin{table}[ht]
\begin{center}
\begin{tabular}{l| l}
                    & \bf QC Classifier  \\ 		
\hline
\hline
Quantum depth $q$       & 15	\\       
Classical depth     	& 1		\\    
Noise variance $N$				& 0.6	\\ 
Training batches 		& 1000	\\       
Batch size      		& 7		\\           
Learning rate	    	& 0.01	\\
Fock-space cutoff\;	    & 11	\\
\hline
Accuracy    & 0.803            
\end{tabular}
\end{center}

\caption{Hyper-parameters used for our quantum state classifier based on the QC transfer learning scheme. The last line reports the corresponding accuracy achieved by the model, simulated on Strawberry Fields with a fixed cutoff in the Fock basis.} \label{tab:tetris}
\end{table}

\begin{figure}[t]
\includegraphics[width=0.48 \textwidth]{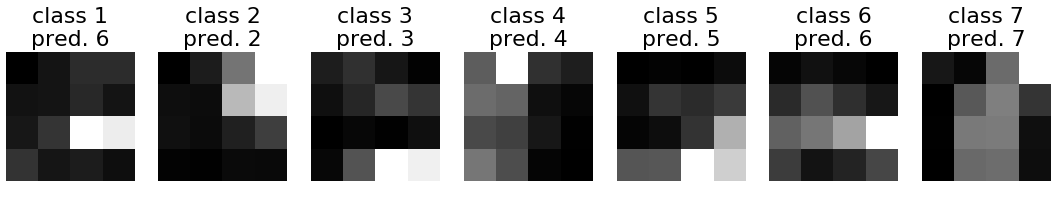}
\caption{Features of a batch of 7 noisy states extracted by the quantum network $A'$ and represented as $4 \times 4$ gray-scale images. The associated classes and the predictions made by our quantum-classical model are reported above each image.}\label{fig:tetris_predictions}
\end{figure}

Finally, the predictions for a sample of 7 noisy states are graphically visualized in Fig.~\ref{fig:tetris_predictions}, where the features extracted by the pre-trained quantum network $A'$ are represented as $4\times4$ gray scale images. The features of Fig.~\ref{fig:tetris_predictions} are quite different from the original tetrominos images shown in Fig.~\ref{fig:tetris}. This due to the truncation of network $A$ and to the presence of input noise. However, as long as the images of Fig.~\ref{fig:tetris_predictions} are distinguishable, this is not a relevant issue since the final classical layer is still able to correctly classify the input states with high accuracy.

\begin{figure}[t]
\includegraphics[width=0.5 \textwidth]{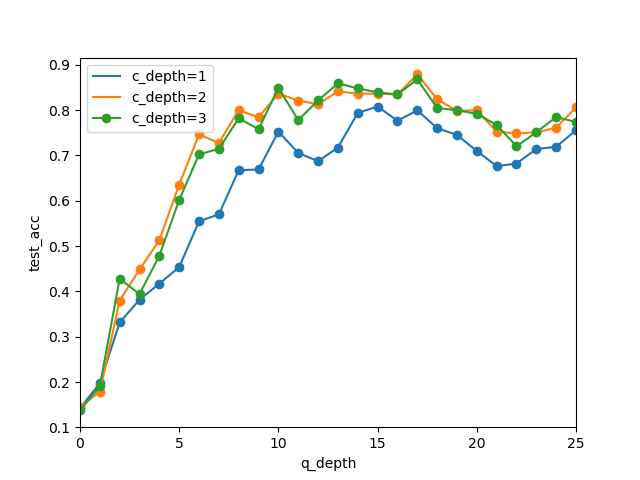}
\caption{Accuracy of the hybrid QC classifier with respect to the quantum depth of the pre-trained network $A'$, evaluated for three different classical networks $B$ of depth $1,2,$ and $3$ respectively. The existence of an intermediate optimal value for the quantum depth is a characteristic signature typical of the transfer learning method.
}\label{fig:tetris_depth}
\end{figure}

We conclude this example with an analysis of the model performance with respect to the values of the quantum and classical depths.
Since the original pre-trained network $A$ has $25$ quantum layers, for the truncated network $A'$ we can choose a quantum depth $q$ within the interval 0-25. For the classical network $B$ we consider the cases of $1,2$ and $3$ layers, corresponding to the models  $L_{16 \rightarrow 7}$, $L_{16 \rightarrow 7} \circ L_{16 \rightarrow 16}$ and  $L_{16 \rightarrow 7} \circ L_{16 \rightarrow 16} \circ L_{16 \rightarrow 16}$, respectively. 

The results are shown in Fig.~\ref{fig:tetris_depth}. By direct inspection we can see that increasing the classical depth is helpful but it saturates the accuracy already after two layers. On the other hand, it is evident that the quantum depth has an optimal value around $q=15$ while for larger values the accuracy is reduced. This is a paradigmatic phenomenon well known in classical transfer learning: better features are usually extracted after removing some of the final layers of $A$. Notice that because of the quantum nature of the system, the quantum state produced by the truncated variational circuit could be entangled and/or not aligned with the measurement basis. So the numerical evidence that the truncation of a quantum network does not always reduce the quality of the measured features, but it can actually be a convenient strategy for transfer learning, is a notable result.

\subsection*{Example 5 -  QQ transfer learning for quantum state classification (Gaussian / non-Gaussian)}

Finally, our last example is a proof-of-principle demonstration of QQ transfer learning. In this case we train an optical network $A$ to classify a particular dataset $D_A$ of Gaussian and non-Gaussian quantum states. Successively, we use it as a pre-trained block for a dataset $D_B$ consisting of Gaussian and non-Gaussian states which are different from those of $D_A$. The pre-trained block is followed by some quantum variational layers that will be trained to classify the quantum states of $D_B$.

Before presenting our model we need to define a continuous-variable single-mode variational layer, the analog of Eq.~\eqref{eq:L_44}. 
We follow the general structure proposed in Ref.\ \cite{killoran2018continuous}:
\begin{align}
\mathcal L =  \Phi \circ \mathcal D \circ \mathcal R \circ \mathcal S \circ \mathcal R, \label{eq:cv_layer}
\end{align}
where $\mathcal R$ is a phase space rotation, $\mathcal S$ is a squeezing operation, $\mathcal D$ is a displacement and $\Phi$ is a cubic phase gate. 
All operations depend on variational parameters and, for sufficiently many layer applications, the model can generate any single-mode unitary operation.
Moreover, by simply removing the last non-Gaussian gate $\Phi$ from \eqref{eq:cv_layer}, we obtain a Gaussian layer which can generate all Gaussian unitary operations.
\begin{figure}[t]
\includegraphics[width=0.45 \textwidth]{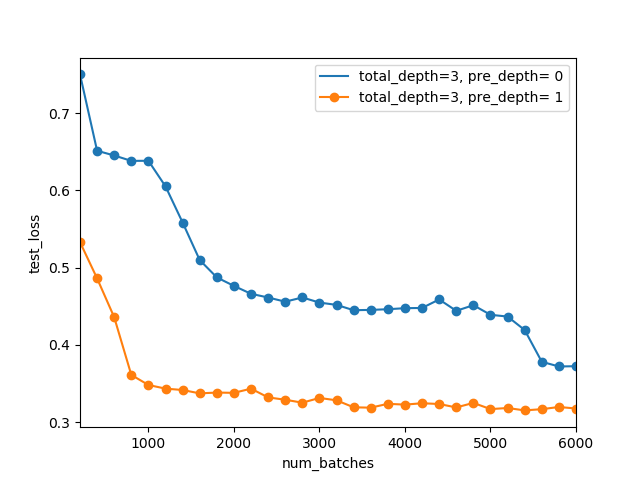}
\includegraphics[width=0.45 \textwidth]{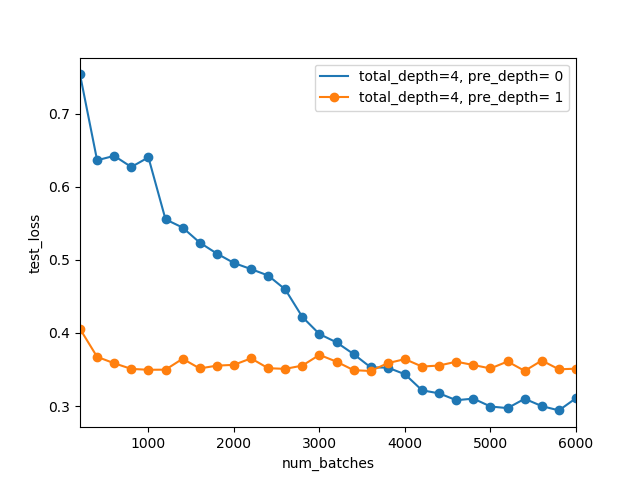}
\caption{Evolution of the loss function (cross entropy) with respect to the number of training iterations. The top plot represents a network of total depth 3 optimized with a QQ transfer learning scheme (orange) compared with a network trained from scratch (blue). In the bottom plot instead the total depth is fixed to 4 layers.} \label{fig:q2q}
\end{figure}

\begin{table}[t]
\begin{center}
\begin{tabular}{l| l}
                    & \bf QQ Classifier  \\ 		
\hline
\hline
Depth of $A'$            & 1 	\\   
Depth of $B$             & 3 	\\  
Training batches 		 & 500	\\       
Batch size      		 & 8		\\           
Learning rate	    	 & 0.01	\\     
Fock-space cutoff\;	     & 15	\\   
\hline
Accuracy    & 0.869            
\end{tabular}
\end{center}
\caption{Hyper-parameters used for our quantum state classifier based on the QQ transfer learning scheme. The last line reports the corresponding accuracy achieved by the model (numerically simulated).} \label{tab:q2q}
\end{table}

We can express the QQ transfer learning model of this example according to the notation introduced in  Section \ref{sec:transfer} and represented in Fig.~\ref{fig:transfer}:
\begin{description}
\item[$D_A$ =] Two classes, 0 and 1, of quantum states generated by two different variational random circuits. States of class 0 are generated by a random single-mode Gaussian layer applied to the vacuum. States of class 1 are generated by a random non-Gaussian layer applied to the vacuum.
\item[$A$ =] Single-mode variational quantum layer followed by an on/off threshold detector.
\item[$T_A$ =] Classification (labels: 0 and 1).
\item[$A'$ =] Network $A$ without the measurement layer.
\item[$D_B$ =] Two classes, 0 and 1, of quantum states. States of class 0 are generated by a random single-mode Gaussian layer applied to the coherent state $|\alpha \rangle$ with $\alpha=1$. States of class 1 are generated by a random Gaussian layer applied to the Fock state $|1\rangle$. 
\item[$B$ =]  Single-mode variational quantum circuit of depth $q$, followed by a on/off threshold detector.
\end{description}

A summary of the hyper-parameters used for defining and training this QQ model is reported in Table \ref{tab:q2q}, together with the associated accuracy.
In Fig.~\ref{fig:q2q} we plot the loss function (cross entropy) of our quantum variational classifier with respect to the number of training iterations. We compare the results obtained with and without the pre-trained layer $A'$ (i.e., with and without transfer learning), for a fixed total depth of 3 or 4 layers.  
It is clear that the QQ transfer learning approach offers a strong advantage in terms of training efficiency. 

For a sufficiently long training time however, the network optimized from scratch achieves the same or better results with respect to the network with a fixed initial layer $A'$. This effect is well known also in the classical setting and it is not surprising: the network trained from scratch is in principle more powerful by construction, because it has more variational parameters. However, there are many practical situations in which the training resources are limited (especially when dealing with real NISQ devices) or in which the dataset $D_B$ is experimentally much more expensive with respect to $D_A$. In all these kind of practically constrained situations, QQ transfer learning could represent a very convenient strategy.

\section{Conclusions}

We have outlined a framework of transfer learning which is applicable to hybrid computational models where variational quantum circuits can be connected to classical neural networks.
With respect to the well-studied classical scenario, in hybrid systems several new and promising opportunities naturally emerge as, for example, the possibility of transferring some pre-acquired knowledge at the classical-quantum interface (CQ and QC transfer learning) or between two quantum networks (QQ transfer learning).
As an additional contribution, we have also introduced the notion of ``dressed quantum circuits", i.e., variational quantum circuits augmented with two trainable classical layers which improve and simplify the data encoding and decoding phases.

Each theoretical idea proposed in this work is supported with a proof-of-concept example, numerically demonstrating the validity of our models for practical applications such as image recognition or quantum state classification.
Particular focus has been dedicated to the CQ transfer learning scheme because of its promising potential with currently available quantum computers. In particular we have used the CQ transfer learning method to successfully classify high resolution images with two real quantum processors (by IBM and Rigetti).

From our theoretical and experimental analysis, we can conclude that transfer learning is a promising approach which can be particularly convenient in the context of near-term quantum devices.

\acknowledgments 

We thank Christian Weedbrook for helpful discussions.
The authors would like to thank Rigetti for access to their resources, Forest \cite{smith2016practical}, QCS and Aspen-4-4Q-A backend. We also acknowledge the use of the IBM Q Experience, Qiskit \cite{abraham2019qiskit} and IBM Q 5 Tenerife v1.0.0 (ibmqx4) backend. 




\bibliography{transfer_learning}

\end{document}